\documentclass[aps,pra,twocolumn,groupedaddress,showpacs,amsmath,amssymb]{revtex4}

\usepackage{amssymb}
\usepackage{amsmath}
\usepackage{graphicx}
\usepackage{amsfonts}
\usepackage{epsfig}
\usepackage{revsymb}
\usepackage{bm}
\usepackage{amsmath}
\usepackage{amssymb}
\usepackage{latexsym}
\usepackage{thumbpdf}
\usepackage{hyperref}

\begin{document}

\title{Comment on "Acoustics of tachyon Fermi gas" \newline
[E. Trojan and G.V. Vlasov, ArXiv: 1103.2276]}

\author{S. N. Burmistrov}

\affiliation{Kurchatov Institute, 123182 Moscow, Russia}



\begin{abstract}
In contrast to Trojan and Vlasov \cite{r1}, it is found that an ideal Fermi gas of
tachyons has a subluminous velocity of sound for any particle density and,
therefore, the causality condition for a tachyon gas holds always true. Also, an
ideal Fermi gas of tachyons never possesses an exotic equation of state with the
pressure exceeding the energy density.
\end{abstract}

\pacs{47.75.+f, 03.30.+p, 04.20.Gz, 05.30.-d}

\maketitle

\par
Recently, Trojan and Vlasov \cite{r1} have reported the acoustic properties of
tachyon Fermi ideal gas at zero temperature. In this study they assert that the
Fermi gas of free tachyons, i.e., particles with the energy spectrum
 $$
\varepsilon _k=\sqrt{k^2-m^2}, \;\;\; (k\geqslant m),
 $$
should be unstable if the particle density $n$ of a gas is below some critical
value $n_T$ because its sound velocity $c_s$ exceeds the speed of light, and thus
the causality principle cannot be satisfied. In addition, authors \cite{r1} have
also arrived at the next assertion that the gas of tachyons may represent an
example of the so-called stiff matter when the pressure $P$ exceeds the energy
density $E$, i.e., $P>E$.
\par
Both the statements are erroneous. The false conclusions result from using
incorrect equation (4) from \cite{r1}
 $$
P=\frac{\gamma}{6\pi ^2}\int\frac{\partial\varepsilon _k}{\partial k}f_kk^3\, dk ,
 $$
resulting from inaccurate integration by parts of the thermodynamical potential
$\Omega$. Instead the authors \cite{r1} must use the starting relation which comes
from the definition \cite{r2} of thermodynamical potential $\Omega =-PV$ leading
to
 $$
P=\gamma T\int\frac{d^3k}{(2\pi )^3}\,\ln\bigl(1+e^{(\mu -\varepsilon _p)/T}\bigr)
,
 $$
where $\mu$ is the chemical potential and $\gamma$ is the degeneracy factor. For
$T=0$, the case of interest in \cite{r1}, one has
 $$
P=\gamma\int\frac{d^3k}{(2\pi )^3}\, (\varepsilon _F-\varepsilon _k)=n\varepsilon
_F-E ,
 $$
where $\varepsilon _F=\mu (T=0)=(k_F^2-m^2)^{1/2}$ is the Fermi energy and $E$ is
the energy density of a gas. Using $E$ from \cite{r1},
 $$
E=\frac{\gamma m^4}{16\pi ^2}\bigl[\bigl(2\beta^3 -\beta ) \sqrt{\beta
^2-1\,}-\ln\bigl(\beta+\sqrt{\beta ^2-1\,}\bigr)\,\bigr] ,
 $$
gives another result for the pressure of a tachyon Fermi ideal gas at $T=0$
 $$
P=\frac{\gamma m^4}{16\pi ^2}\bigl[\bigl(\frac{2}{3}\beta^3 +\beta
-\,\frac{8}{3}\bigr)\sqrt{\beta ^2-1\,}+\ln\bigl(\beta+\sqrt{\beta
^2-1\,}\bigr)\,\bigr].
 $$
In addition, the above expression agrees completely with the determination of the
pressure as a derivative of the total energy $EV$ over gas volume $V$ with the
opposite sign
 $$
P=-\partial (EV)/\partial V=n\partial E/\partial n \, - E .
 $$
\par
The square of the sound velocity $c_s$ is then given by
 $$
c_s^2=\partial E/\partial P= \frac{1}{3}\frac{\beta ^2+\beta +1}{\beta ^2+\beta}\,
,
 $$
where $\beta=k_F/m\geqslant 1$ is a ratio of the Fermi momentum to the tachyon
mass. Thus, unlike \cite{r1}, the sound velocity, changing between $1/\sqrt{2}$
and $1/\sqrt{3}$, is always smaller than the speed of light and the causality
principle $c_s<1$ does not break down for any density of ideal tachyon gas.
\par
As it concerns the ratio of pressure to energy density
 $$
\frac{P}{E}=\frac{1}{3}\frac{\bigl(2\beta^3 +3\beta -8\bigr)\sqrt{\beta
^2-1\,}+3\ln\bigl(\beta+\sqrt{\beta ^2-1\,}\bigr)}{\bigl(2\beta^3 -\beta )
\sqrt{\beta ^2-1\,}-\ln\bigl(\beta+\sqrt{\beta ^2-1\,}\bigr)},
 $$
it varies from 1/2 in the low density limit to 1/3 for the high density limit. As
a result, the hypothesis \cite{r1} that the low density gas of free tachyon can
represent an example of the so-called absolute stiff  or hyperstiff matter with
the equation of state $P\geqslant E$ proves to be invalid as well.
\par
In \cite{r3} an attempt is made to resolve a confusion of Ref.~\cite{r1} with the
well-established thermodynamic relations by using the following energy spectrum of
a tachyon
\begin{eqnarray*}
\varepsilon _k=\left\{
\begin{array}{cc}
0, & k< m
\\
\sqrt{k^2-m^2}, & k\geqslant m .
\end{array}
\right.
\end{eqnarray*}
However, such energy spectrum does not satisfy the relativity principle that the
square of four-vector $(\varepsilon ,\,\bm{k})$ must be an identical scalar within
the whole region of permissible $\bm{k}$, i.e.,
 $$
\varepsilon ^2_{\bm{k}}-\bm{k}^2=\text{const}.
 $$
Obviously, this condition breaks down for the spectrum proposed.
\par
In conclusion, the arguments in \cite{r1,r3} to consider a tachyon ideal gas at
sufficiently low density either as an unstable $c_s\geqslant 1$ matter or as an
exotic $P\geqslant E$ matter have no theoretical grounds.

\end{document}